**When the future alters the present – how Discrete Dynamical Systems replicate images**

Sugata Mitra

Tataha Kim Laboratory
NIIT University,
NH8 Delhi-Jaipur Highway,
Neemrana,
District Alwar,
Rajasthan 301705,
India

Email: sugata.mitra@niituniversity.in

**Abstract**

Agents affected by their own future states in a one-dimensional discrete dynamical system (1-DDS) can replicate two-dimensional images. It is shown that such replication requires a toroidal spacetime and three rules are needed to calculate the number of iterations required for exact replication. It is argued that retrocausal updation used by 1-DDS can replicate any n-dimensional digital object. It is shown that the way iterations reach a final image are different for randomly generated images and non-random, *meaningful* images. Two instances of real-world events that seem to imply such retrocausal replication are discussed.

1. **Introduction and background**

This paper is about agents that can look ahead into their future. Images or bitmaps can be fractally replicated by *time manipulated one-dimensional cellular automata* [1]. The present work examines the nature, meaning and physical basis for such replication.

Instead of modified cellular automata, a vertical array of agents forming a one-dimensional discrete dynamical system (1-DDS) were used. Each agent can be in one of two states, 1 and 0, represented by black and white. Each agent's current state is updated by the current states of its two neighbors, above and below, and by its own state in the future. Since its own future state affects its present state, the system is retrocausal. The 1-DDS can update over a fixed number of time steps, t=1 to w, and then return to t=1. Each such cycle is an iteration.

If a bitmap is placed in the future of this 1-DDS, the agents will react to the bits of the image and change in accordance with the updation rules provided. It is known that a modified Wolfram rule 150 will replicate the image in $2^n$ iterations [1].

In this paper, it was found that:

1. Iterations have a physical basis if the image being replicated is considered a surface in toroidal spacetime.

2. The progression of successive iterations towards replicating the final image shows differences for random images when compared with non-random, *meaningful*, images.

3. The relationship between the number of iterations required to replicate an object, $I_{min}$, and the 'size' of the object is a step function of the width of the bitmap, but with exceptions if the number of agents in the 1-DDS is an integral power of 2. This corrects an earlier inadequate explanation [1].

In addition to these findings, the replication of n-dimensional objects and the implications of such replication are discussed.

## 2. Description of the experimental process

The replicator used consists of a retrocausal one-dimensional discrete dynamical system [2] made up of agents [3]. The system consists of a line of discrete agents arrayed in a specific geometry (4). Each agent can be in one of k finite states, and each agent's state is updated on every time step according to a deterministic rule based on the values of the neighboring agents and the future state of the agent being updated. In the computational experiments described below, we have worked with a 1-DDS that is, a single line of agents, placed vertically, each in one of two states – black=1 or white=0.

We have used the following two-state model having a neighborhood of radius 1:

Number of states: k =2

Number of neighbors above and below: r = 1

Retrocausal Update Rule (RUR): Agent $_{t+1}$ = (Above$_t$+Agent$_{t+1}$+Below$_t$) mod 2

This rule implies that every agent's future state at time t+1 exists, and affects the agent's state at time t.

This update rule is a modification of the usual:

Causal Update Rule (CUR): Agent $_{t+1}$ = (Above$_t$+Agent$_t$+Below$_t$) mod 2

CUR, is called Rule 150 according to Wolfram's numbering system [5] and is quite well known [1].

Since our retrocausal update rule involves the future states of agents in a 1-DDS, we can populate that future with a group of states with value 1. This could be an image. If we do so and iterate the 1-DDS over this image using RUR, we observe image replication as expected. If we had instead used CUR for updation, the image would have been erased by the advancing 1-DDS.

## 3. Necessity for a Toroidal Spacetime

Is it possible to have a physically consistent system that allows for such retrocausal iterations of a 1-DDS?

To answer this question, a black-and-white (binary) image of Leonardo Da Vinci was replicated using a retrocausal 1-DDS as described above. The results are shown in figure 1.

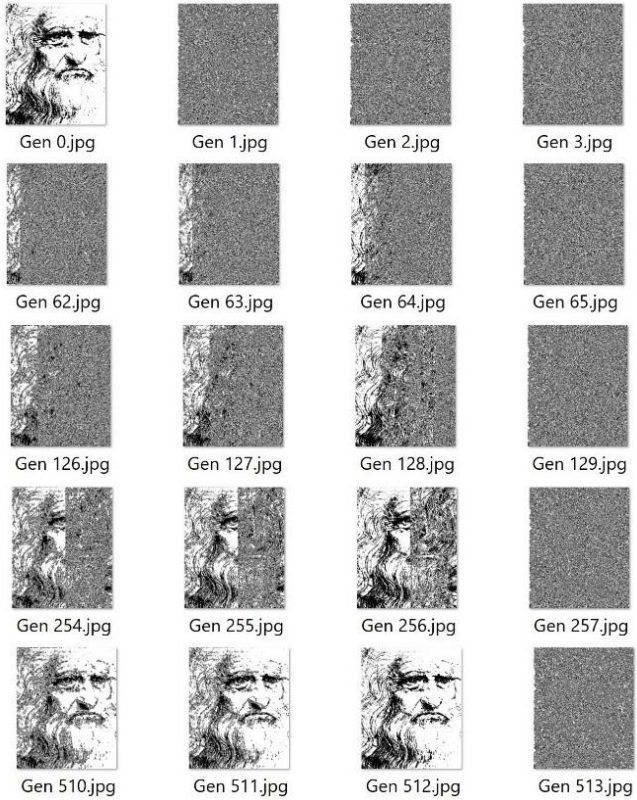

Figure 1. Image replicated by a retrocausal 1-DDS over 512 iterations.

The image used has 480x585 pixels, and contains approximately 25% black pixels. It was exactly replicated in 512 iterations. For each iteration, how closely the iterated image matched the original was calculated. A plot of the percent match (henceforth called *replication efficiency*) against iteration number is shown in figure 2.

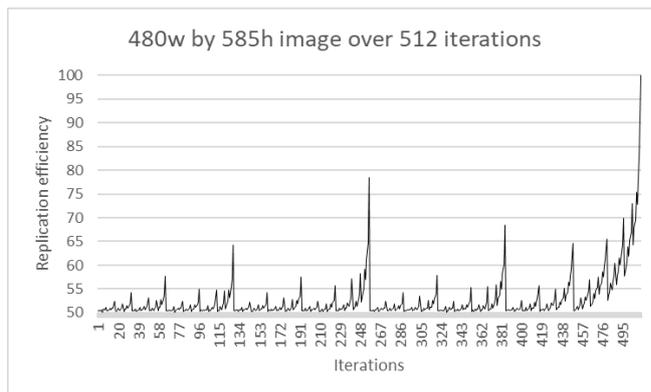

Figure 2. How progressive iterations approach the original image.

Figure 2 shows that the replication efficiency reaches 100% at the 512$^{th}$ iteration. Iterated images get closer to matching with the original image, with peaks in replication efficiency every $2^n$ iterations. Exact match, i.e., 100% is achieved at n=9, the 512$^{th}$ iteration.

Each iteration, some shown in figure 1, involves application of update rule, RUR, until the end of the width of the picture, column 480 in this case, is reached. After this, the next iteration begins at column 1. It is important to realize that the column numbers are units of time and the width of this image represents time. Since our DDS is one-dimensional, the only spatial dimension it has is its height which is actually the number of agents in the system, 585 in this case. The vertical height axis is referred to as y, and the horizontal width axis as t.

If it is assumed that the t-axis is closed, that is, its value returns to zero after 480, a basis for why iterations restart from zero after reaching the maximum width of the image is established.

The update rule involves both neighbors of the agent being updated. This will cause problems at both ends of the y-axis. This problem can be resolved by joining the ends of the y-axis. In other words, the y-axis is closed, and its value returns to zero after 585. This provides a consistent basis for applying RUR along the y-axis.

The shape of a spacetime formed by a closed spatial axis, y, and a closed temporal axis, t, is a two-dimensional, orthogonal torus as shown in figure 3. The torus shape removes problems related to edges – whether it is the maximum and minimum values on the y- axis, or the future and past on the t-axis. The application of such toroidal spacetime to cellular automata is not common in the literature, however, there have been some attempts [6].

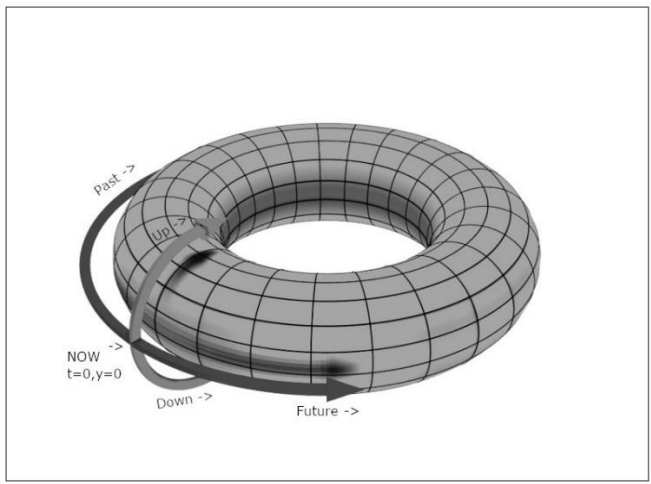

**Figure 3. Toroidal spacetime with one temporal and one spatial axis.**

An image or a group of pixels (bitmap) wrapped such that it fits exactly onto a torus such as the one shown in figure 3, would enable update rule RUR to apply over the time axis in the direction of the future until it covers the circumference of the time axis and returns to zero. This would complete a single iteration and the process would repeat, thereby replicating the image or bitmap repeatedly as shown in figures 1 and 2. We would have a physical analogue for the process of replication, provided we assume that any bitmap creates a toroidal spacetime around which it is draped.

It is also important to note that the bitmaps we are working with are two-dimensional, while the DDS we are using to replicate the bitmaps with are one-dimensional. A 1-DDS is essentially a binary string, a column of agents in one of two states. The update rule RUR produces a new binary string at each time step. These strings when laid side by side produces a two-dimensional image. In other words, we are interpreting the time axis as an additional spatial axis to visualize an "image". What we are referring to as an "image" is a memory of past states of the column of agents, frozen as a spatial axis.

4. **Number of iterations required to replicate images**

The minimum number of iterations required for exact replication is called $I_{min}$, as multiples of $I_{min}$ iterations will also produce exact replications. We know from the experiments above that $I_{min} = 2^n$, where n is an integer. In the earlier paper [1], it was suggested that $I_{min}$ is related to the area of the image, and two special cases were discussed. However, no general method was suggested for calculating $I_{min}$ for an image of width (w) and height (h). In this paper, we will derive such a method.

We need to know if $I_{min}$ is related to the height of the image on the y-axis, the width of the image on the t-axis, or both. To do this, we will need equivalent images of different heights and widths. We generated images consisting of random black and white dots to conduct these experiments.

First, it was checked whether an image consisting of random dots would be equivalent to a non-random ("meaningful") image. The image of Leonardo in figure 1 consists of 480x585 pixels (black or white dots) with 25% black pixels. A random bitmap was created by distributing the same number of black pixels randomly over an image of the same dimensions. This was replicated using the RUR on a 1-DDS. Figure 4 shows the replication efficiency between iterations 400-512.

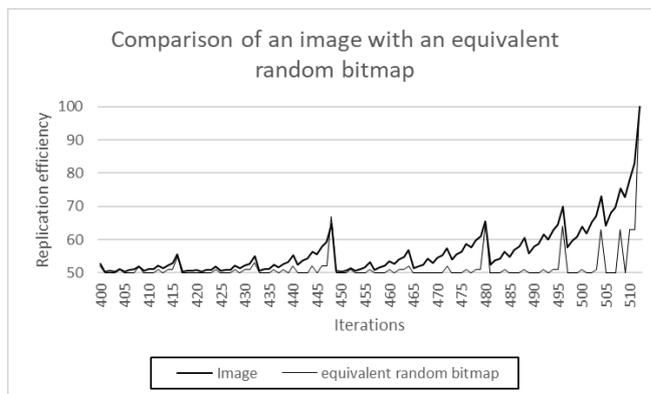

Figure 4. Comparison of replication efficiencies for random and non-random images.

The match percent data for the Leonardo image of Figure 1 showed a 76% correlation with that for the equivalent random bitmap used for Figure 4. More importantly perhaps, the peaks in both graphs are at identical powers of 2. We concluded that while the matching progresses somewhat differently for the two images, the number of iterations required for exact replication are identical. This was noticed for several images compared with equivalent random bitmaps. This difference between random and non-random, *meaningful*, images seems to indicate that height, width, and pixel density are not the only determinants that shape the replication efficiency graph. The distribution of pixels, whether random or non-random, affects the shape of the graphs in figure 4, even though they don't affect the value of $I_{min}$. *Meaningful* images are non-random. Therefore, the separation between the graphs for meaningful images and equivalent random images can be an indicator of *meaning*.

However, for the purpose of determining the relationship between $I_{min}$ and the height and width of an image, random images were used for the experiments below.

Experiment 1: The y-axis height(h) of the image was varied from 1 to 256 in steps of 1, while keeping the width (w) constant at 64. At each step a random image was generated and iterated using RUR until exact replication was achieved. The results are shown in figure 5.

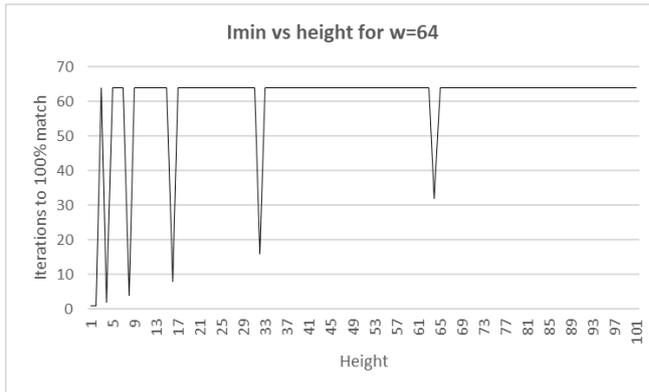

Figure 5. Iterations needed to reach 100% replication for images with varying heights and fixed width

Figure 5 shows that an image of width 64 units will require 64 iterations to replicate irrespective of its height, except when h is an integral power of 2 and is 64 or less. In our example, h meets these conditions for values 2,4,8,16,32 and 64. At these values, $I_{min}$ =h/2. Several other experiments confirmed these findings.

In general if h<=w and h=$2^n$, then $I_{min}$ =$2^{n-1}$, n being an integer. When h>w, $I_{min} = 2^{\lceil \log_2 w+1 \rceil}$ and independent of h. Figure 5 shows that $I_{min}$ is determined by the value of w, except when h is an exact integral power of 2.

Experiment 2: The t-axis width(w) of the image was varied from 1 to 256 in steps of 1, while keeping the height (h) constant at 64. At each step a random image was generated and iterated using RUR until exact replication was achieved. The results are shown in figure 6.

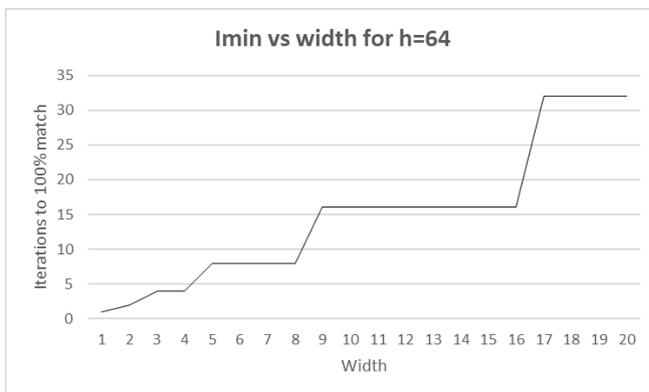

Figure 6. Iterations needed to reach 100% replication for images with varying widths and fixed height

Figure 6 shows that for images with widths less than height, an image of fixed height (64 units in this case) will require $2^n$ iterations for an exact match (where n is an integer) such that $2^{n-1}$<=w<$2^n$. For example in Figure 6, for values of w<h, $I_{min}$=$2^4$ for widths where $2^3$<=w<$2^4$.

In general, $I_{min} = 2^{\lceil \log_2 w+1 \rceil}$ . Except when h is an integral power of 2.

$I_{min}$ is a step function of width, except when the height is an integral power of 2. To check this, heights of 63,64,65 and 128 were tested. Figure 7 shows the results. At h=63 where $2^5$<h<=$2^6$ and at h=65 where $2^6$<h<=$2^7$, $I_{min}$ changes as a step function of width. However, at h=64, $I_{min}$ becomes independent of width and remains constant at 32 which is $2^5$. Similarly, at h=128, $I_{min}$ becomes independent of width and remains constant at 64 which is $2^6$. The individual graphs for each height is shown on the top of Figure 7, followed by the combined graph.

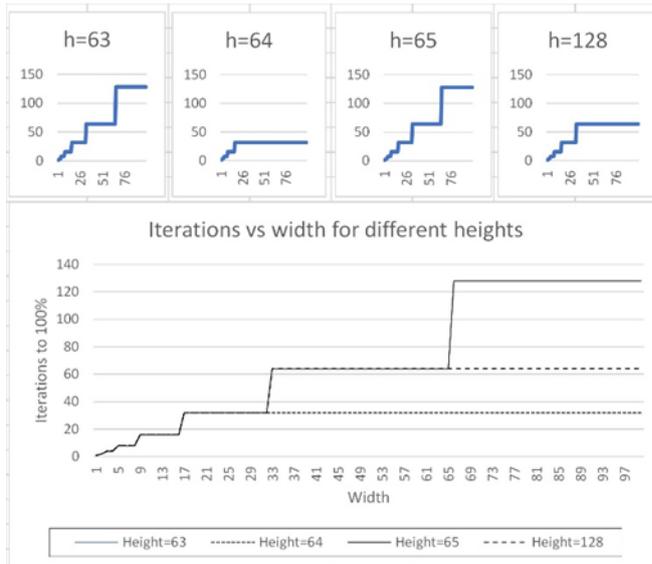

Figure 7. Relationship of I$_{min}$ with width for different heights

Next, widths that were 63, 64 and 65 were tested. These are shown in Figure 8. As expected, at widths of 63 and 64 where $2^5<=w<2^6$, I$_{min}$ is $2^6$. At a width of 65, where $2^6<=w<2^7$, I$_{min}$ jumps to $2^7$.

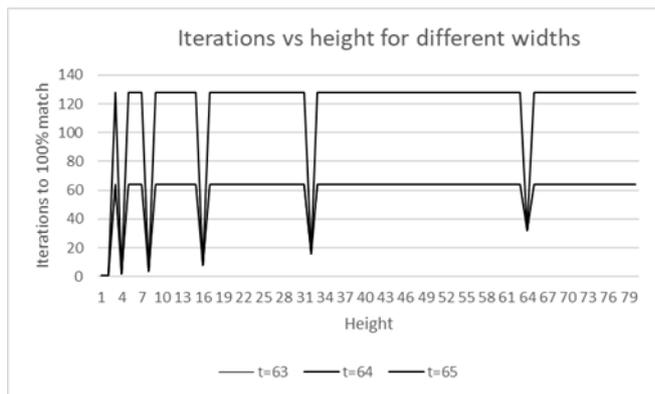

Figure 8. Relationship of I$_{min}$ with height for different widths

We can generalize these results into the following rules:

For a bitmap of dimensions w and h:

1. $I_{min} = 2^{\lceil \log_2 w + 1 \rceil}$, for all values of w when h is not a positive integral power of 2.
2. $I_{min} = 2^{\lceil \log_2 w + 1 \rceil}$, for all values of w when h is a positive integral power of 2 and h=>w.
3. $I_{min} = 2^{\lceil \log_2 h - 1 \rceil}$, for all values of h that are positive integral powers of 2 and h<=w.

I$_{min}$ is always determined by the integral powers of 2 on the time axis (w), except when the number of agents making up the 1-DDS, the y axis (h) value, equals a positive integral power of 2. Hence, I$_{min}$ can be made independent of w by using the right number of agents (h) in a 1-DDS.

We can use these rules to predict the value of I$_{min}$ for a bitmap of any dimensions.

5. Discussion

A. Any digital object can be replicated by a retrocausal 1-DDS

In the work described above, two-dimensional bitmap images were replicated. Such replication is possible for any digitally represented object because all files used to represent objects are a sequence of bytes that can be represented as a binary string of bits. Any such binary string can be 'folded' into a rectangle of sides h and w. Such rectangular bitmaps can be replicated in $2^n$ iterations by 1-DDS using the retrocausal update rule RUR in a toroidal spacetime. The replicated bitmap can then be 'unfolded' into the original binary string and finally, converted to the sequence of bytes that constitute the original digital object.

**B. Extension to n-dimensional bitmaps**

Would DDS with more than one spatial dimension also replicate digital objects if iterated over the time axis using an appropriate updation rule? One way to answer this question would be through experiments with multi-dimensional DDS. Such an attempt was made in [1] to replicate images using a two-dimensional cellular automaton and using Conway's 'Game of Life' updation rules [7], modified for retrocausality. This was not an equivalent experiment to our experiments with 1-DDS. The experiment reported in [1] consisted of replicating a two-dimensional image using a 2-DDS. An equivalent experiment should have consisted of a 2-DDS in a toroidal spacetime of two spatial dimensions and a time dimension, encountering a three-dimensional object.

It is possible to design such an experiment; however, it is not necessary to do so. A two-dimensional image made of 'on' and 'off' pixels can be represented as a one-dimensional binary string if we know how to fold the string at the right points to make a two-dimensional matrix (in this case 'matrix', 'bitmap' and 'image' are used interchangeably). Such a matrix if laid out in columns over a time axis can be replicated by RUR. We know from the experiments above that a two-dimensional matrix of pixels draped over a toroidal spacetime (Figure 3) can be repeatedly replicated using the updation rule RUR. We can extend this thinking to three-dimensional objects. A group of pixels in three-dimensional space can be laid out as two-dimensional 'slices' over a time axis. Each image can be further reduced to a one-dimensional binary string. This string folded into a rectangle can be replicated by RUR as described in various instances above. Doing this over all the slices of the original object would, in effect, replicate the three-dimensional object. We can also extend this reasoning to objects with higher dimensions.

In general, n-dimensional bitmaps can be successively reduced until they are represented by a binary, one-dimensional string. All we need to know is how to 'fold' the string to create the original n-dimensional object. This one-dimensional string can then be converted to a rectangle where one side is a spatial dimension and the other side is time. A 1-DDS and the retrocausal updation rule, RUR, will then replicate this rectangle in an appropriate toroidal spacetime. In effect, the original n-dimensional object would be repeatedly replicated.

Retrocausal updation using 1-DDS can modify objects in n-dimensions that can be 'frozen' in n+1 dimension to produce hyper objects. Beyond n=2, we do not have the ability to visualize such objects.

**C. Effect of iterations on meaning**

Random images and meaningful non-random images show differences in replication curves as seen in section 4. RUR based Iterations seem to either destroy or create meaning. An example is shown in figure 9.

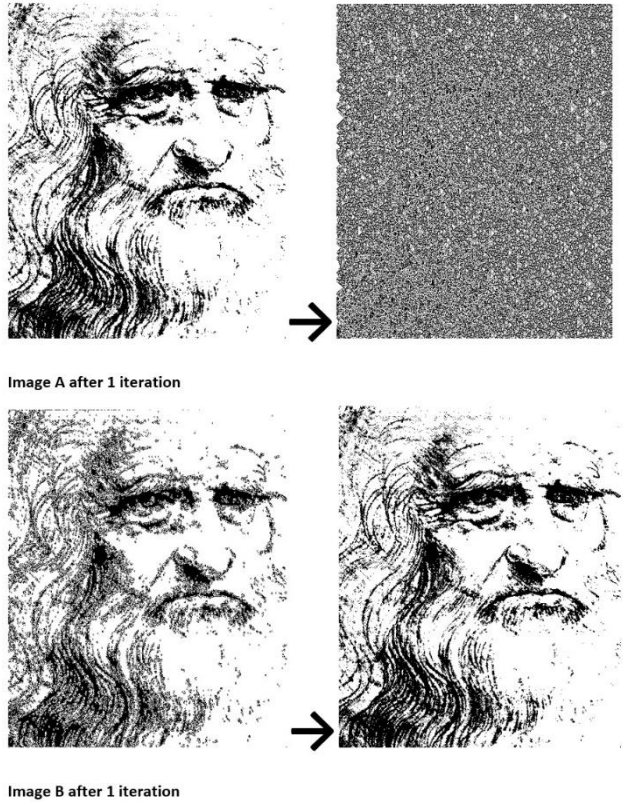

Figure 9. The effect of a single iteration on two, nearly identical, images

Images A and B in figure 9 are different by about 17%, and nearly identical visually. Image A is the original image while image B is the result of 511 iterations on image A. A single iteration destroys all meaning in image A while a single iteration on image B produces an exact replication of the original image. Each pixel in these images are states of agents in a 1-DDS as described before. Figure 9 shows that the states of a few agents in a 1-DDS can affect the results of an iteration to exceptionally large extents.

The progression of iterations over time are irreversible. If an agent has a value of 1 at time t, it is not possible to determine what its value at time t-1 was. This is because the updation rule is such that the same state can be produced by different combinations of the states of itself and its upper and lower neighbors in the previous time step. Hence, RUR is deterministic in the forward, past to present to future, direction of time, but not the other way about.

Further, it is not possible to derive the image from its replication efficiency graph. For example, figure 4 indicates the existence of a non-random image, but there is no method for finding what that image was. Further research may uncover such a method.

The process of replication of non-random images appears to be one where iterations create and destroy meaning in cycles, until the final image is reached.

Meaning in binary strings is not understood clearly although some attempts have been made (e.g., [8]).

**D. Applicability to other cyclical systems**

Figure 10 shows Wolfram's diagram (Figure 11,[5]) of the number of occurrences of the 1-bit in the binary representation of an integer n. Figure 10 also shows the iterations 400-512 of the Leonardo image referred to earlier. While two similar graphs do not necessarily imply similar causes, we cannot rule out the possibility. In this case, integral powers of 2 play a role in both graphs.

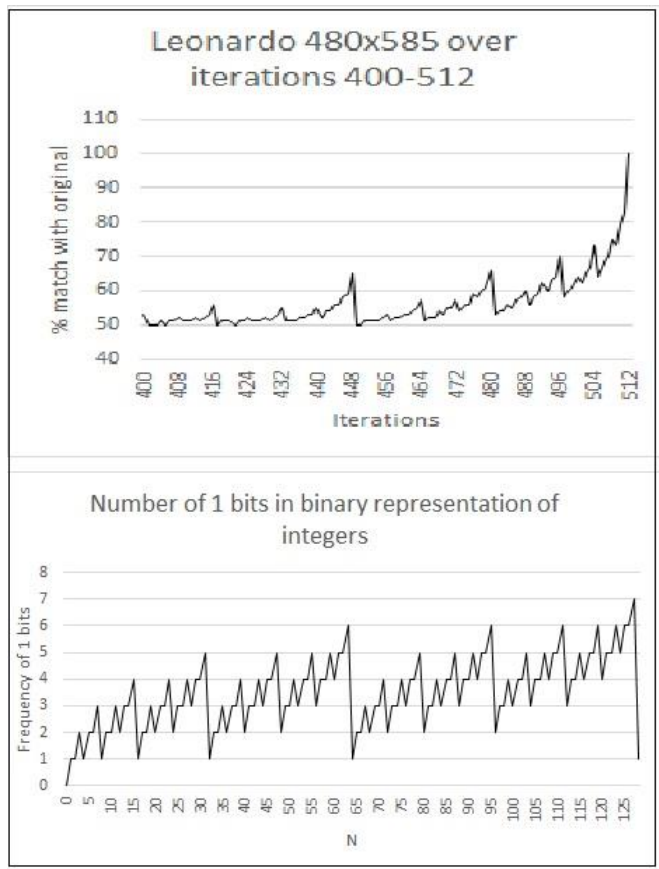

Figure 10. A section of figure 1 (top) with a reconstruction of Wolfram's graph below.

The worldwide COVID-19 virus pandemic [9] shows cyclical patters that can be compared with 1-DDS iteration graphs. The growth of infections show peaks and troughs over time, heading upwards. This is typical of the behavior of replication efficiency in bitmap replication by retrocausal agents in a 1-DDS. Figure 11 shows such a comparison.

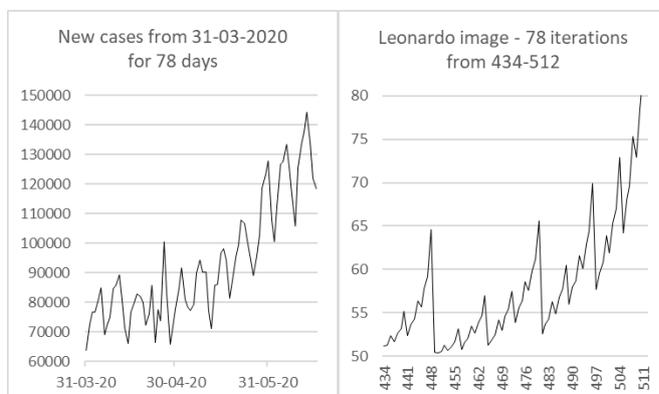

Figure 11. COVID-19 infections worldwide compared to a replication graph of 1-DDS

The same mechanisms may not be involved in these two instances, however, that possibility cannot be ruled out. There is a 75% correlation between the graphs in Figure 11. The correlation between the Covid data and an equivalent random bitmap is 20%. This is shown in Figure 12.

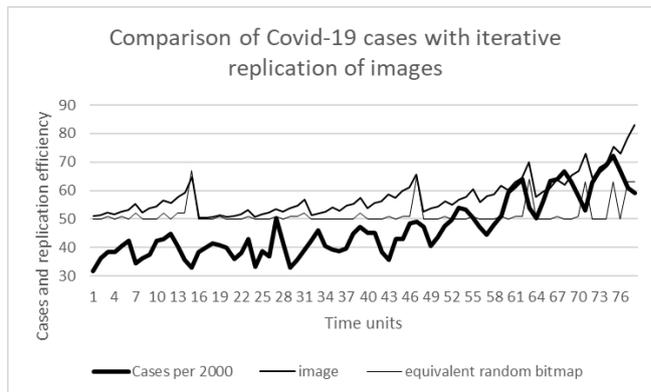

Figure 12. Worldwide Covid cases between 31st May and 16th June 2020, compared with replication efficiency of an image and an equivalent random bitmap.

Virus infection involves neighborhood interactions to determine future states. If the mechanism of the spread of the Covid virus is like that of a 1-DDS replicating an image, Figure 11 tells us that image is non-random. If this is so, the $I_{min}$ values for replication can be made constant and independent of time by changing the number of agents involved, as shown in section 4.

It could be suspected that the graph of infections in figures 11 and 12 suggest a retrocausal 1-DDS iterating over time to replicate an *image*, or a pattern that we may not be able to comprehend.

**Funding**: This research did not receive any specific grant from funding agencies in the public, commercial, or not-for-profit sectors.

**Declaration of interest**: None

**References**

1. Mitra, Sugata and Kumar, Sujai. (2005). 'Fractal Replication in Time Manipulated One-Dimensional Cellular Automata', Complex Systems, Vol. 16 (3). Mitra and Kumar
2. An introduction to discrete dynamical systems, Internet: "Math Insight" https://mathinsight.org/discrete_dynamical_system_introduction
3. Bonabeau, Eric (2002). Agent-based modeling: Methods and techniques for simulating human systems, PNAS vol. 99 suppl. 3, pp7280–7287. https://www.pnas.org/content/99/suppl_3/7280 .
4. Shiffman, D. (2012). Chapter 7, The Nature of Code – Simulating Natural Systems with Code. https://www.amazon.com/Nature-Code-Simulating-Natural-Processing/dp/0985930802
5. Wolfram, S. (1983). "Statistical Mechanics of Cellular Automata," Reviews of Modern Physics, 55 601–644. http://lattice.ifsc.usp.br/~lattice/oldlattice/artigo-wolfram-cellular-autom.pdf
6. Rouhaud, J. (2000). Cellular automata and consumer behaviour, European Journal of Economic and Social Systems 14 N° 1 37-52. https://ejess.edpsciences.org/articles/ejess/pdf/2000/01/rouhaud.pdf?access=ok
7. Gardner, M. (1983). Wheels, Life and other Mathematical Amusements (W. H. Freeman, New York.

8. Mitra, Sugata.(2002) 'Meaning in Binary Strings', Introduction to Multimedia Systems, Academic Press, pp. 151. Also downloadable from https://7551850c-18fa-45d6-98bc-8e2829cc3850.filesusr.com/ugd/369ee5_d5ddac04b3e14a37939ed8b1f8f521f3.pdf
9. Corona Virus Data (2020), from ourworldindata.org https://ourworldindata.org/coronavirus-source-data